\documentclass[british,conference,final,twocolumn,11pt]{IEEEtran}
\usepackage[T1]{fontenc}
\usepackage[latin9]{inputenc}
\usepackage[a4paper]{geometry}
\usepackage{verbatim}
\usepackage{float}
\usepackage{amsmath}
\usepackage{graphicx}

\makeatletter

\providecommand{\tabularnewline}{\\}
\floatstyle{ruled}
\newfloat{algorithm}{tbp}{loa}
\providecommand{\algorithmname}{Algorithm}
\floatname{algorithm}{\protect\algorithmname}

\usepackage{algorithmic}

\makeatother

\usepackage{babel}
\begin{document}

\title{Low-Complexity Iterative Sinusoidal Parameter Estimation}

\author{Jean-Marc Valin$^{\star\diamond}$, Daniel V. Smith$^{\dagger}$,
Christopher Montgomery$^{\ddagger\diamond}$, Timothy B. Terriberry$^{\diamond}$\\
$^{\star}$CSIRO ICT Centre, Australia\\
$^{\dagger}$CSIRO Tasmanian ICT Centre, Australia\\
$^{\ddagger}$RedHat Inc., USA\\
$^{\diamond}$Xiph.Org Foundation\\
\{jean-marc.valin,daniel.v.smith\}@csiro.au\\
\{xiphmont,tterribe\}@xiph.org}
\maketitle
\begin{abstract}
\footnotetext{\copyright 2007 IEEE.  Personal use of this material is permitted. Permission from IEEE must be obtained for all other uses, in any current or future media, including reprinting/republishing this material for advertising or promotional purposes, creating new collective works, for resale or redistribution to servers or lists, or reuse of any copyrighted component of this work in other works.}
Sinusoidal parameter estimation is a computationally-intensive task,
which can pose problems for real-time implementations. In this paper,
we propose a low-complexity iterative method for estimating sinusoidal
parameters that is based on the linearisation of the model around
an initial frequency estimate. We show that for $N$ sinusoids in
a frame of length $L$, the proposed method has a complexity of $O\left(LN\right)$,
which is significantly less than the matching pursuits method. Furthermore,
the proposed method is shown to be more accurate than the matching
pursuits and time frequency reassignment methods in our experiments. 
\end{abstract}

\section{Introduction}

The sinusoidal model is increasingly being used in signal processing
applications such as speech synthesis \cite{Stylianou2001}, speech
coding \cite{Hedelin1983}, and audio coding \cite{LevineThesis}.
Estimating the model parameters often represents a significant fraction
of the overall complexity of these applications. However, many real-time
applications require a very low-complexity estimation algorithm.

This paper proposes a new procedure based on the linearisation of
the model around an initial frequency estimate. Parameters are optimised
using an iterative method with fast convergence. We show that for
typical configurations, it is over 20 times less complex than matching
pursuits \cite{mallat93matching}. 

We start by introducing sinusoidal modelling and prior art in Section
\ref{sec:Sinusoidal-Modelling}. Section \ref{sec:Frequency-Estimation}
discusses frequency estimation and our proposed linearisation. In
Section \ref{sec:Iterative-Solver}, we present a low-complexity iterative
solver for estimating sinusoidal parameters. Results are presented
in Section \ref{sec:Results-And-Discussion} with a discussion and
Section \ref{sec:Conclusion} concludes this paper.

\section{Sinusoidal Parameter Estimation\label{sec:Sinusoidal-Modelling}}

A general sinusoidal model that considers both amplitude and frequency
modulation can be used to approximate a signal $\tilde{x}\left(t\right)$
as:
\begin{equation}
\tilde{x}\left(t\right)=\sum_{k=1}^{N}A_{k}\left(t\right)\cos\left(\int_{0}^{t}\omega_{k}\left(t\right)dt+\phi_{k}\right)\ ,\label{eq:general-sinusoidal}
\end{equation}
where $A_{k}\left(t\right)$ is the time-varying amplitude, $\omega_{k}\left(t\right)$
is the time-varying frequency and $\phi_{k}$ is the initial phase.
The model in (\ref{eq:general-sinusoidal}) has limited practical
use because it is very complex and has an infinite number of ways
to approximate $x\left(t\right)$. Using discrete time $n$ and normalised
frequencies $\theta_{k}$ over a finite window $h\left(n\right)$
yields a simpler model:
\begin{equation}
\tilde{x}\left(n\right)=h\left(n\right)\sum_{k=1}^{N}\left(A_{k}+A_{k}^{'}n\right)\cos\left(\theta_{k}n+\phi_{k}\right)\ ,\label{eq:4-term-general}
\end{equation}
where $A^{'}$ is the first time derivative of the amplitude, or even
\begin{equation}
x\left(n\right)=h\left(n\right)\sum_{k=1}^{N}A_{k}\cos\left(\theta_{k}n+\phi_{k}\right)\label{eq:3-term-general}
\end{equation}
if we do not want to model amplitude variation within a frame. Although
simpler, the models in (\ref{eq:4-term-general}) and (\ref{eq:3-term-general})
are still difficult to estimate because they involve a non-linear
optimisation problem.

Several methods exist for estimating sinusoidal parameters. The standard
DFT over a rectangular window is limited by both frequency leaking
(sidelobes from the rectangular window) and its poor frequency resolution
equal to $\pi/L\ rad/s$ for a frame of length $L$. 

By defining an over-complete dictionary of sinusoidal bases, matching
pursuits methods \cite{mallat93matching} make it possible to increase
the resolution arbitrarily, while allowing for a window in its basis
functions. However, being a greedy algorithm, matching pursuits behaves
sub-optimally when the basis functions used are not orthogonal \cite{vos50hqc},
which is usually the case for sinusoids of arbitrary frequency over
a finite window length. The orthogonality problem of matching pursuits
can be mainly overcome by further non-linear optimisation as in \cite{vos50hqc}.
However this requires a significant increase in complexity (such as
$O\left(N^{4}\right)$ terms). 

Another approach is the time frequency (TF) reassignment method, which
can be used to improve estimates of frequency localisation within
various forms of TF representations \cite{Auger1995}, including spectrograms
\cite{Auger1995,Plante1998}. In the case of the spectrogram, phase
information from the short time Fourier transform (STFT) is exploited
to move energy away from the centre of the frequency bin $(t,w)$
to the centre of gravity of the spectral distribution \cite{Auger1995}.
Hence, this approach can be used to reduce the inaccuracy of frequency
estimation in a quantised TF representation that is reliant upon the
temporal resolution of the window. A drawback to this approach is
that it is not well suited to noisy signal conditions, as energy becomes
concentrated in noise dominated regions \cite{Plante1998}.

Other work, such as \cite{Stylianou2001,Dhaes2004} focuses on the
estimation of sinusoidal partials in harmonic signals. While these
methods generally have a low complexity, they are not applicable to
non-harmonic signals.

\section{Linearised Model\label{sec:Frequency-Estimation}}

As another way of obtaining accurate frequency estimation, we propose
rewriting the sinusoidal model in (\ref{eq:4-term-general}) as
\begin{multline}
\tilde{x}\left(n\right)=h\left(n\right)\sum_{k=1}^{N}\left(A_{k}+nA_{k}^{'}\right)\cdot\\
\cos\left(\left(\theta_{k}+\Delta\theta_{k}\right)n+\phi_{k}\right)\ ,\label{eq:5-term-general}
\end{multline}
where $\theta_{k}$ is an initial estimate of the frequencies and
$\Delta\theta_{k}$ is an unknown correction to the initial estimate.
When both the amplitude modulation parameter $A_{k}^{'}$ and the
frequency correction $\Delta\theta_{k}$ are small, we show in Appendix
\ref{sec:Linearisation} that (\ref{eq:5-term-general}) can be linearised
as the sum of four basis functions
\begin{multline}
\tilde{x}\left(n\right)=h\left(n\right)\sum_{k=1}^{N}c_{k}\cos\theta_{k}n+s_{k}\sin\theta_{k}n\\
+d_{k}n\cos\theta_{k}n+t_{k}n\sin\theta_{k}n\ ,\label{eq:least-square-prob}
\end{multline}
with
\begin{align}
c_{k}= & A_{k}\cos\phi_{k}\ ,\label{eq:lin_ak_param}\\
s_{k}= & -A_{k}\sin\phi_{k}\ ,\\
d_{k}= & A_{k}^{'}\cos\phi_{k}-A_{k}\Delta\theta_{k}\sin\phi_{k}\ ,\\
t_{k}= & -A_{k}^{'}\sin\phi_{k}-A_{k}\Delta\theta_{k}\cos\phi_{k}\ .\label{eq:lin_dk_param}
\end{align}

From now on, unless otherwise noted, bold uppercase symbols ($\mathbf{A}$)
denote matrices, bold lower case symbols ($\mathbf{a}_{i}$) denote
the columns of the matrix and italic symbols ($a_{i,j}$) denote the
elements of the matrix. We can express (\ref{eq:least-square-prob})
in matrix form as
\begin{align}
\tilde{\mathbf{x}}= & \mathbf{A}\mathbf{w}\ ,\\
\mathbf{A}= & \left[\mathbf{A}^{c},\mathbf{A}^{s},\mathbf{A}^{d},\mathbf{A}^{t}\right]\ ,\\
\mathbf{w}= & \left[\mathbf{c},\mathbf{s},\mathbf{d},\mathbf{t}\right]^{T}\ .
\end{align}
where the basis components $\mathbf{A}^{c}$, $\mathbf{A}^{s}$, $\mathbf{A}^{s}$
and $\mathbf{A}^{t}$ are defined as
\begin{align}
a_{n,k}^{c}= & h\left(n\right)\cos\theta_{k}n\ ,\label{eq:basis_fn1}\\
a_{n,k}^{s}= & h\left(n\right)\sin\theta_{k}n\ ,\\
a_{n,k}^{d}= & h\left(n\right)n\cos\theta_{k}n\ ,\\
a_{n,k}^{t}= & h\left(n\right)n\sin\theta_{k}n\ .\label{eq:basis_fn4}
\end{align}
The best fit can then be obtained through a least-square optimisation,
by posing

\begin{equation}
\frac{\partial}{\partial\mathbf{w}}\left\Vert \mathbf{A}\mathbf{w}-\mathbf{x}_{h}\right\Vert ^{2}=0\ ,
\end{equation}
where $\mathbf{x}_{h}$ is the windowed input signal. This leads to
the well known solution
\begin{equation}
\mathbf{w}=\left(\mathbf{A}^{T}\mathbf{A}\right)^{-1}\mathbf{A}^{T}\mathbf{x}_{h}\ .\label{eq:explicit-least-square}
\end{equation}

Once all linear parameters in (\ref{eq:least-square-prob}) are found,
the real sinusoidal parameters can be retrieved by solving the system
(\ref{eq:lin_ak_param})-(\ref{eq:lin_dk_param}):
\begin{align}
A_{k} & =\sqrt{c_{k}^{2}+s_{k}^{2}}\ ,\\
\phi_{k} & =\arg\left(c_{k}-\jmath s_{k}\right)\ ,\\
A_{k}^{'} & =\frac{d_{k}c_{k}+s_{k}t_{k}}{A_{k}}\ ,\\
\Delta\theta_{k} & =\frac{d_{k}s_{k}-t_{k}c_{k}}{A_{k}^{2}}\ .
\end{align}

\section{Iterative Solver\label{sec:Iterative-Solver}}

Though it is far less computationally demanding than a classic non-linear
solver, solving the linear system (\ref{eq:explicit-least-square})
still requires a great amount of computation. In \cite{Dhaes2004},
a method was proposed to reduce that complexity from $O\left(LN^{2}\right)$
to $O\left(N\log N\right)$, but only for harmonic signals. In this
paper, we propose an $O\left(LN\right)$ solution without the restriction
to harmonic signals.

Our method uses an iterative solution based on the assumption that
matrix $A$ is close to orthogonal, so that
\begin{equation}
\left(\mathbf{A}^{T}\mathbf{A}\right)^{-1}\approx\mathrm{diag}\left\{ \frac{1}{\mathbf{a}_{1}^{T}\mathbf{a}_{1}},\ldots,\frac{1}{\mathbf{a}_{N}^{T}\mathbf{a}_{N}}\right\} =\boldsymbol{\Phi}\ .
\end{equation}
That way, an initial estimate can be computed as
\begin{equation}
\mathbf{w}^{(0)}=\boldsymbol{\Phi}^{-1}\mathbf{A}^{T}\mathbf{x}_{h}\label{eq:Jacobi-initial}
\end{equation}
and then refined as
\begin{align}
\mathbf{w}^{(i+1)} & =\mathbf{w}^{(i)}+\boldsymbol{\Phi}^{-1}\mathbf{A}^{T}\left(\mathbf{x}_{h}-\tilde{\mathbf{x}}^{(i)}\right)\nonumber \\
 & =\mathbf{w}^{(i)}+\boldsymbol{\Phi}^{-1}\mathbf{A}^{T}\left(\mathbf{x}_{h}-\mathbf{A}\mathbf{w}^{(i)}\right)\ .\label{eq:Jacobi-update}
\end{align}

It turns out that the iterative method described in (\ref{eq:Jacobi-initial})-(\ref{eq:Jacobi-update})
is strictly equivalent to the Jacobi iterative method. The complexity
of the algorithm is reduced to $O(LMN)$, where $M$ is the number
of iterations required for acceptable convergence. Unfortunately,
while the Jacobi method is generally stable for most matrices $\mathbf{A}$
obtained in practice, convergence is not guaranteed and depends on
the actual frequencies $\theta_{k}$.

\subsection{Gauss-Seidel Method}

An alternate to the Jacobi method is the Gauss-Seidel method, which
has the main advantage that it is guaranteed to converge in this case
because the matrix $\mathbf{A}^{T}\mathbf{A}$ is a symmetric, positive
definite matrix \cite{saad2003ims}. Because the columns of $\mathbf{A}$
are usually nearly orthogonal, $\mathbf{A}^{T}\mathbf{A}$ is strongly
diagonally dominant and the Gauss-Seidel method converges quickly.
The linear system can be expressed as 
\begin{equation}
\mathbf{R}\mathbf{w}=\mathbf{b}\ ,
\end{equation}
where
\begin{align}
\mathbf{R} & =\mathbf{A}^{T}\mathbf{A}\ ,\\
\mathbf{b} & =\mathbf{A}^{T}\mathbf{x}_{h}\ .
\end{align}
If we assume that matrix $\mathbf{A}$ has been pre-normalised ($\mathbf{a}_{k}^{T}\mathbf{a}_{k}=1,\forall k$),
the Gauss-Seidel algorithm is expressed as
\begin{align}
w_{k}^{\left(i+1\right)}= & b_{k}-\sum_{j<k}r_{k,j}w_{j}^{\left(i+1\right)}-\sum_{j>k}r_{k,j}w_{j}^{\left(i\right)}\nonumber \\
= & \mathbf{a}_{k}^{T}\mathbf{x}_{h}-\sum_{j<k}\mathbf{a}_{k}^{T}\mathbf{a}_{j}w_{j}^{\left(i+1\right)}-\sum_{j>k}\mathbf{a}_{k}^{T}\mathbf{a}_{j}w_{j}^{\left(i\right)}\nonumber \\
= & w_{k}^{\left(i\right)}+\mathbf{a}_{k}^{T}\mathbf{x}_{h}-\sum_{j<k}\mathbf{a}_{k}^{T}\mathbf{a}_{j}w_{j}^{\left(i+1\right)}\nonumber \\
 & -\sum_{j\geq k}\mathbf{a}_{k}^{T}\mathbf{a}_{j}w_{j}^{\left(i\right)}\nonumber \\
= & w_{k}^{\left(i\right)}+\mathbf{a}_{k}^{T}\mathbf{x}_{h}-\mathbf{a}_{k}^{T}\left(\mathbf{A}\tilde{\mathbf{w}_{k}}^{\left(i+1\right)}\right)\nonumber \\
= & w_{k}^{\left(i\right)}+\mathbf{a}_{k}^{T}\left(\mathbf{x}_{h}-\mathbf{A}\tilde{\mathbf{w}_{k}}^{\left(i+1\right)}\right)\ ,\label{eq:GS1-final}
\end{align}
where
\begin{multline}
\tilde{\mathbf{w}_{k}}^{\left(i+1\right)}=\left[w_{0}^{\left(i+1\right)},\ldots,w_{k-1}^{\left(i+1\right)},\right.\\
\left.w_{k}^{\left(i\right)},\ldots,w_{N-1}^{\left(i\right)}\right]^{T}\ .
\end{multline}
We can further simplify the computation of (\ref{eq:GS1-final}) by
noting that only one element of $\tilde{\mathbf{w}_{k}}^{\left(i+1\right)}$
changes for each step. We thus have
\begin{equation}
w_{k}^{\left(i+1\right)}=w_{k}^{\left(i\right)}+\mathbf{a}_{k}^{T}\mathbf{e}_{k}^{\left(i+1\right)}\ ,
\end{equation}
where $\mathbf{e}_{k}^{\left(i+1\right)}$ is the current error on
the approximation and is computed recursively as
\begin{equation}
\mathbf{e}_{k}^{\left(i+1\right)}=\left\{ \begin{array}{ll}
\begin{aligned} & \mathbf{e}_{k-1}^{\left(i+1\right)}-\\
 & \quad\left(w_{k-1}^{\left(i+1\right)}-w_{k-1}^{\left(i\right)}\right)\mathbf{a}_{k-1}
\end{aligned}
 & ,\ k\neq0\\
\mathbf{e}_{N}^{\left(i\right)} & ,\ k=0
\end{array}\right.\ .
\end{equation}
The resulting computation is summarised in Algorithm \ref{alg:Improved-iterative-algorithm}.
If there is only one iteration, then algorithm \ref{alg:Improved-iterative-algorithm}
is equivalent to a simplified version of the matching pursuits algorithm
where the atoms (frequency of the sinusoids) have been pre-defined
before the search. From this point of view, the proposed method relaxes
the orthogonality assumption made by the matching pursuits method.

The main difference with the Jacobi method is the Gauss-Seidel method
includes partial updates of the error term after each extracted sinusoid.
The convergence also follows intuitively from the fact that each individual
step is an exact projection that is guaranteed to decrease the current
error $\mathbf{e}$ --- or at worst leave it constant if the optimal
solution has been reached. Also, because the error term is updated
after each component $k$, placing the highest-energy terms first
speeds up the optimisation. For this reason, we first include the
$\cos\theta_{k}n$ and the $\sin\theta_{k}n$ terms, followed by the
$n\cos\theta_{k}n$ and the $n\sin\theta_{k}n$ terms. We have found
that this usually reduces the number of iterations required for convergence.
The resulting algorithm typically converges in half as many iterations
as alternative conjugate gradient techniques, such as LSQR \cite{PaigeSaunders1982},
which cannot take advantage of the diagonal dominance of the system.

\begin{algorithm}
\begin{algorithmic}

\STATE Compute basis functions (\ref{eq:basis_fn1})-(\ref{eq:basis_fn4}).

\STATE $\mathbf{w}^{(0)}\leftarrow\mathbf{0}$

\STATE $\mathbf{e}\leftarrow\mathbf{x}_{h}$

\FORALL{iteration $i$=1\ldots M}

\FORALL{sinusoid component $k=1\ldots 4N$}

\STATE $\Delta w_{k}^{(i)}\leftarrow\mathbf{a}_{k}^{T}\mathbf{e}$

\STATE $\mathbf{e}\leftarrow\mathbf{e}-\mathbf{a}_{k}\Delta w_{k}^{(i)}$

\STATE $w_{k}^{(i)}\leftarrow w_{k}^{(i-1)}+\Delta w_{k}^{(i)}$

\ENDFOR

\ENDFOR

\FORALL{sinusoid $k=1\ldots N$}

\STATE$A_{k}\leftarrow\sqrt{c_{k}^{2}+s_{k}^{2}}$

\STATE$\phi_{k}\leftarrow\arg\left(c_{k}-\jmath s_{k}\right)$

\STATE$A_{k}^{'}\leftarrow\frac{d_{k}c_{k}+s_{k}t_{k}}{A_{k}}$

\STATE$\Delta\theta_{k}\leftarrow\frac{d_{k}s_{k}-t_{k}c_{k}}{A_{k}^{2}}$

\ENDFOR

\end{algorithmic}

\caption{Iterative linear optimisation\label{alg:Improved-iterative-algorithm}}
\end{algorithm}

If in (\ref{eq:basis_fn1})-(\ref{eq:basis_fn4}) we (arbitrarily)
choose $n=0$ to lie in the centre of the frame (between sample $L/2$
and sample $L/2+1$ if $L$ is even), the $\mathbf{a}_{k}^{c}$ and
$\mathbf{a}_{k}^{t}$ vectors all have even symmetry, while $\mathbf{a}_{k}^{s}$
and $\mathbf{a}_{k}^{d}$ all have odd symmetry. This leads to the
following orthogonality properties:
\begin{align}
\left\langle \mathbf{a}_{k}^{c},\mathbf{a}_{k}^{s}\right\rangle  & =0\ ,\label{eq:ortho1}\\
\left\langle \mathbf{a}_{k}^{c},\mathbf{a}_{k}^{d}\right\rangle  & =0\ ,\\
\left\langle \mathbf{a}_{k}^{t},\mathbf{a}_{k}^{s}\right\rangle  & =0\ ,\\
\left\langle \mathbf{a}_{k}^{t},\mathbf{a}_{k}^{d}\right\rangle  & =0\ .\label{eq:ortho4}
\end{align}
Because the even and odd bases are orthogonal to each other, we can
optimise them separately as
\begin{align}
\left[\mathbf{c},\mathbf{t}\right]^{T}= & \left(\mathbf{A}^{even}{}^{T}\mathbf{A}^{even}\right)^{-1}\mathbf{A}^{even}{}^{T}\mathbf{x}\ ,\\
\left[\mathbf{d},\mathbf{s}\right]^{T}= & \left(\mathbf{A}^{odd}{}^{T}\mathbf{A}^{odd}\right)^{-1}\mathbf{A}^{odd}{}^{T}\mathbf{x}\ ,\\
\mathbf{A}^{even}= & \left[\mathbf{A}^{c},\mathbf{A}^{t}\right]\ ,\\
\mathbf{A}^{odd}= & \left[\mathbf{A}^{d},\mathbf{A}^{s}\right]\ .
\end{align}
Not only does the orthogonality accelerate convergence, but it allows
us to split the error $\mathbf{e}$ into half-length even and odd
components, reducing the complexity of each iteration by half.

\subsection{Non-Linear Optimisation}

If the initial frequency estimates $\theta_{k}^{0}$ are close to
the real frequency of the sinusoids $\theta_{k}$, then the error
caused by the linearisation (\ref{eq:least-square-prob}) is very
small. In this case, Algorithm \ref{alg:Improved-iterative-algorithm}
should result in a value of $\theta_{k}^{0}+\Delta\theta_{k}$ that
is even closer to the real frequencies. However, if the initial estimates
deviate significantly from the real values, then it may be useful
to restart the optimisation from the new frequency estimates. Repeating
the operation several times, we obtain a non-linear iterative solver
for $A_{k}$, $\theta_{k}$, $A_{k}^{'}$ and $\phi_{k}$. 

We have found that it is not necessary to wait for Algorithm \ref{alg:Improved-iterative-algorithm}
to converge before updating the frequencies $\theta_{k}$. We can
let both the linear part and the non-linear part of the solution run
simultaneously. To do that, we must first subtract the solution of
the previous iteration before restarting the linear optimisation. 

\begin{algorithm}
\begin{algorithmic}

\STATE$\forall k,\ \theta_{k}=initial\ frequency\ estimate$

\STATE$\forall k,\ [A_{k},\phi_{k},A_{k}^{'}]\leftarrow0$

\STATE $\mathbf{w}^{(0)}\leftarrow\mathbf{0}$

\STATE $\mathbf{e}\leftarrow\mathbf{x}_{h}$

\FORALL{non-linear iteration $i$=1\ldots M}

\FORALL{sinusoid $k$}

\STATE$c_{k}\leftarrow A_{k}\cos\phi_{k}$

\STATE$s_{k}\leftarrow-A_{k}\sin\phi_{k}$

\STATE$d_{k}\leftarrow A_{k}^{'}\cos\phi_{k}$

\STATE$t_{k}\leftarrow-A_{k}^{'}\sin\phi_{k}$

\ENDFOR

\STATE$\mathbf{e}\leftarrow\mathbf{x}-\mathbf{A}\mathbf{w}^{(i-1)}$
(result of the last iteration with updated frequency)

\FORALL{sinusoid component $k=1\ldots 4N$}

\STATE $\Delta w_{k}^{(i)}\leftarrow\mathbf{a}_{k}^{T}\mathbf{e}$

\STATE $\mathbf{e}\leftarrow\mathbf{e}-\mathbf{a}_{k}\Delta w_{k}^{(i)}$

\STATE $w_{k}^{(i)}\leftarrow w_{k}^{(i-1)}+\Delta w_{k}^{(i)}$

\ENDFOR

\FORALL{sinusoid $k=1\ldots N$}

\STATE$A_{k}\leftarrow\sqrt{c_{k}^{2}+s_{k}^{2}}$

\STATE$\phi_{k}\leftarrow\arg\left(c_{k}-\jmath s_{k}\right)$

\STATE$A_{k}^{'}\leftarrow\frac{d_{k}c_{k}+s_{k}t_{k}}{A_{k}}$

\STATE$\theta_{k}\leftarrow\theta_{k}+\frac{d_{k}s_{k}-t_{k}c_{k}}{A_{k}^{2}}$

\ENDFOR

\ENDFOR

\end{algorithmic}

\caption{Non-linear iterative optimisation\label{alg:Non-linear-iterative-optimisation}}
\end{algorithm}

The non-linear method we propose is detailed in Algorithm \ref{alg:Non-linear-iterative-optimisation}
and shares some similarities with the Gauss-Newton method \cite{GaussNewton}.
However, because the reparametrisation in (\ref{eq:lin_ak_param})-(\ref{eq:lin_dk_param})
allows updates to $A_{k}$, $A^{'}$ and $\phi_{k}$ to be incorporated
into the linear model immediately when solving the normal equations,
convergence is greatly improved compared to a standard Gauss-Newton
iteration in the original parameters. Just like Algorithm \ref{alg:Improved-iterative-algorithm},
it is possible to reduce the complexity of Algorithm \ref{alg:Non-linear-iterative-optimisation}
by half by taking advantage of the even-odd symmetry of the basis
functions.

\section{Results And Discussion\label{sec:Results-And-Discussion}}

In this section, we characterise the proposed algorithm and compare
it to other sinusoidal parameter estimation algorithms. We attempt
to make the comparison as fair as possible despite the fact that the
methods we are comparing do not have exactly the same assumptions
or output. Both the linear and the non-linear versions of the proposed
algorithm are evaluated. For all algorithms, we use a \emph{sine window}
\begin{equation}
h(n)=\cos\pi\frac{n-\left(L+1\right)/2}{L}\ ,
\end{equation}
so that the result of applying the window to both the input signal
$\mathbf{x}$ and the basis functions $\mathbf{a}_{k}$ is equivalent
to a Hanning analysis window. Unless otherwise noted, we use a frame
length $L=256$.

\subsection{Convergence}

\begin{figure}
\includegraphics[width=1\columnwidth]{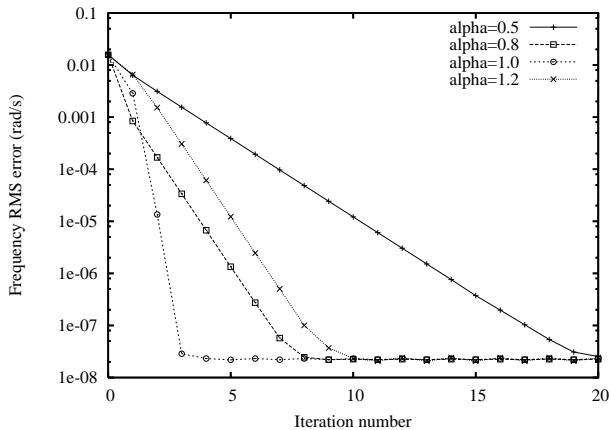}

\caption{Convergence of the non-linear optimisation procedure for various values
of $\alpha$. For $\alpha=1$, convergence is achieved in only 3 iterations.
The floor at $2\times10^{-8}\:rad/s$ is due to the finite machine
precision.\label{fig:Convergence-of-nonlinear-optimisation}}
\end{figure}

We first consider the case of a single amplitude-modulated sinusoid
of normalised angular frequency $\theta=0.1\pi$. We start with an
initial frequency estimate of $\theta=0.095\pi$, which corresponds
to an error of slightly more than one period over the 256-sample frames
we use. The non-linear optimisation Algorithm \ref{alg:Non-linear-iterative-optimisation}
is applied with different values of $\alpha$. The convergence speed
in Figure \ref{fig:Convergence-of-nonlinear-optimisation} shows that
for $\alpha=1$, convergence becomes much faster than for other values
of $\alpha$. This provides a strong indication that the convergence
of the algorithm is super-linear, although we give no formal proof.

\subsection{Chirps}

We measure the frequency estimation accuracy and the energy of the
residual signal for known signals. We use a synthetic signal that
is the sum of five chirps with white Gaussian noise. The chirps have
linear frequency variations starting at $0.05,\:0.1,\:0.15,\:0.2,\:0.25\:rad/s$
and ending at $2.0,\:2.2,\:2.4,\:2.6,\:2.8\:rad/s$, respectively.
The relative amplitudes of the chirps are 0 dB, -3 dB, -6 dB, -9 dB,
and -12 dB. The following algorithms are considered:
\begin{itemize}
\item Time frequency reassignment (\textbf{TFR}),
\item Matching pursuits (32x over-sampled dictionary) (\textbf{MP}),
\item Proposed algorithm with linear optimisation (\textbf{linear}), and
\item Proposed algorithm with non-linear optimisation (\textbf{non-linear}).
\end{itemize}
The time frequency reassignment method is implemented as in \cite{Auger1995}.
The matching pursuits algorithm uses a dictionary of non-modulated
sinusoids with a resolution of $\pi/8192$. We also compare to the
theoretical resolution obtained from the picking the highest peaks
in the DFT. To make sure that algorithms are compared fairly, all
algorithms are constrained to frequencies within one DFT bin of the
real frequency, i.e. there are no outliers. 

\begin{figure}
\includegraphics[width=1\columnwidth]{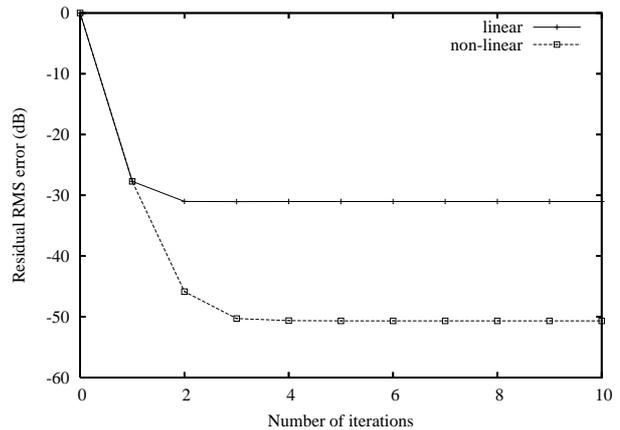}

\caption{Reconstruction RMS error as a function of the number of iterations
in clean conditions (linear vs. non-linear)\label{fig:Reconstruction-convergence}}
\end{figure}

Fig. \ref{fig:Reconstruction-convergence} shows the RMS energy of
the residual ($\tilde{\mathbf{x}}-\mathbf{x}_{h}$) as a function
of the number of iterations for both the linear optimisation and the
non-linear optimisation. The linear version converges after only 2
iterations, while the non-linear version requires 3 iterations. We
use 3 iterations for both methods in the experiments that follow.

\begin{figure}
\includegraphics[width=1\columnwidth]{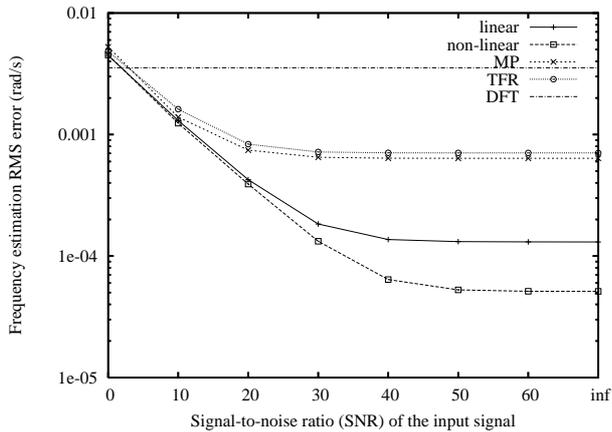}

\caption{Frequency RMS estimation error as a function of the SNR.\label{fig:Frequency-RMS-error}}
\end{figure}

Fig.\ref{fig:Frequency-RMS-error} shows the frequency RMS estimation
error as a function of the SNR for each of the four algorithms. At
very low SNR, all algorithms perform similarly. However, as the SNR
increases above 20 dB, matching pursuits stops improving. This is
likely due to the fact that the frequencies are not orthogonal, which
makes its greedy approach sub-optimal. Both the proposed linear and
non-linear approaches provide roughly the same accuracy up to 30 dB,
after which the non-linear approach provides superior performance.
For this scenario, the only limitation of the non-linear algorithm
at infinite SNR is the fact that it does not account for frequency
modulation within a frame.

\begin{figure}
\includegraphics[width=1\columnwidth]{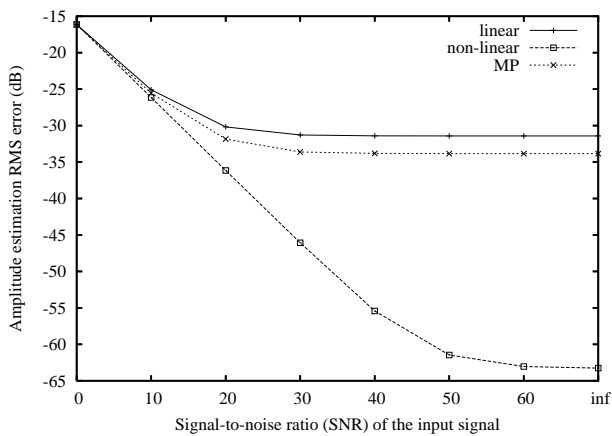}

\caption{Amplitude RMS estimation error as a function of the SNR.\label{fig:Amplitude-RMS-error}}
\end{figure}

The amplitude estimation error is shown in Fig. \ref{fig:Amplitude-RMS-error}.
Although the behaviour of the amplitude error is similar to that of
the frequency estimation error, the difference between the linear
and the non-linear optimisation algorithms is accentuated. The time
frequency reassignment algorithm is not included in the comparison
because it does not estimate amplitude.

\begin{figure}
\includegraphics[width=1\columnwidth]{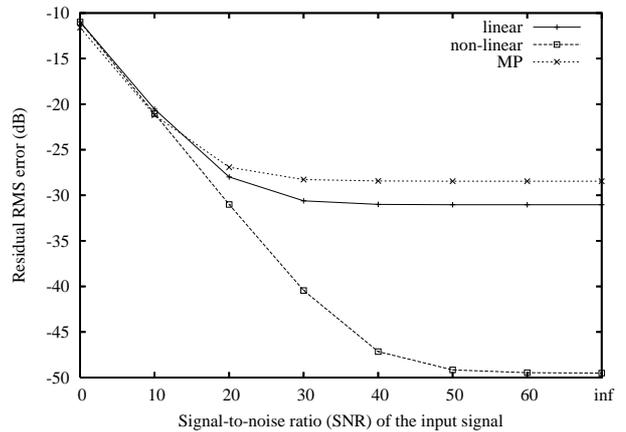}

\caption{Reconstruction RMS error as a function of the SNR (the input noise
is not considered in the error).\label{fig:Reconstruction-RMS-error}}
\end{figure}

Fig. \ref{fig:Reconstruction-RMS-error} compares the reconstruction
error for all algorithms, except the time frequency reassignment method,
which cannot estimate the amplitude and thus cannot provide a reconstructed
signal. The reconstruction error is computed based on the non-noisy
version of the chirps. We observe performances similar to the ones
in Fig. \ref{fig:Frequency-RMS-error}, with the notable exception
that when it comes to reconstruction, the non-linear optimisation
is able to fit the data much more efficiently than the linear optimisation
at high SNR. 

We also observe that the performance of our algorithm is slightly
worse than that of matching pursuits at low SNR. This can be explained
by some slight over-fitting due to the fact that the proposed algorithm
also includes an amplitude modulation term. The difference disappears
if the amplitude modulation term is forced to zero.

Overall, we observe from the experiment on chirps that our proposed
non-linear algorithm clearly out-performs both matching pursuits and
time frequency reassignment. The linear version has overall slightly
better performance than the other methods, although it does not perform
as well as non-linear optimisation. In all cases (Fig. \ref{fig:Frequency-RMS-error}
to Fig. \ref{fig:Reconstruction-RMS-error}), all the algorithms compared
behave similarly. Their error at low SNR is similar and the slope
of the improvement is the same. What differentiates the algorithms
is how far they improve with SNR before reaching a plateau.

\subsection{Audio}

We apply our proposed algorithm to a 90-second collage of diverse
music clips sampled at 48 kHz, including percussive, musical, and
amusical content. In this case, we cannot compare to the matching
pursuits algorithm because the lack of \emph{ground truth} prevents
us from forcing a common set of initial sinusoid frequencies. We select
the initial frequency estimates required for the proposed algorithm
using peaks in the standard DFT.

\begin{figure}

\includegraphics[width=1\columnwidth]{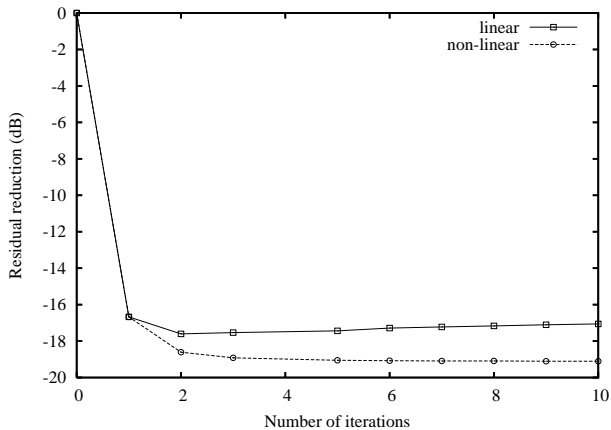}

\caption{Reduction in residual energy as a function of the number of iterations.\label{fig:Audio-Reduction-in-residual}}
\end{figure}

The energy of the residual is plotted as a function of the number
of iterations in Fig. \ref{fig:Audio-Reduction-in-residual}. Both
algorithms converge quickly and we can see that the linear optimisation
only requires 2 iterations, while the non-linear optimisation requires
3 iterations. 

\subsection{Algorithm complexity}

In this section, we compare the complexity of the proposed algorithms
to that of other similar algorithms. For the sake of simplicity, we
discard some terms that are deemed negligible, e.g., we discard $O\left(LN\right)$
terms when $O\left(LN^{2}\right)$ terms are present.

In Algorithm \ref{alg:Improved-iterative-algorithm}, we can see that
each iteration requires $8LN$ multiplications and $8LN$ additions.
Additionally, computation of the $4N$ basis functions $\mathbf{a}_{k}$
prior to the optimisation requires $LN$ additions and $3LN$ multiplications.
It is possible to further reduce the complexity of each iteration
by taking advantage of the fact that all of our basis functions have
either even or odd symmetry. By decomposing the residual into half-length
even and odd components, only one of these components needs to be
updated for a given basis function. This reduces the complexity of
each iteration in Algorithm \ref{alg:Improved-iterative-algorithm}
by half without changing the result. The complexity of each iteration
is thus $4LN$ multiplications and $4LN$ additions. For $M$ iterations,
this amounts to a total of $\left(8M+5\right)LN$ operations per frame. 

The complexity of the proposed non-linear optimisation algorithm (Algorithm
\ref{alg:Non-linear-iterative-optimisation}) is similar to that of
the linear version, with two notable exceptions. First, because the
frequency is changing for every iteration, the basis functions need
to be re-computed for every iteration. Second, when starting a new
iteration, the residual must be updated using the new basis functions.
The total complexity is thus $\left(17M-4\right)LN$ operations per
frame (for a single iteration, the linear and non-linear versions
are strictly equivalent).

As a comparison a simple matching pursuits algorithm that does not
consider modulation requires $4LN^{2}P$ operations per frame, where
$P$ is the oversampling factor (i.e. increase over the standard DFT
resolution). If a fast (FFT-based) implementation of the matching
pursuits algorithm \cite{vos50hqc} is used, then the complexity is
reduced to $5/2LNP\log_{2}LP$. 

Table \ref{tab:Complexity-comparison} summarises the complexity of
several algorithms. Because the algorithms have different dependencies
on all the parameters, we also consider the total complexity in Mflops
for real-time estimation of sinusoids in a \emph{typical} scenario,
where we have
\begin{itemize}
\item frame length: $L=256$,
\item number of sinusoids: $N=20$,
\item oversampling: $P=32$ (matching pursuits only),
\item number of iterations: $M=2$ (linear), $M=3$ (non-linear),
\item sampling rate: 48 kHz,
\item frame offset: 192 samples (25\% overlap).
\end{itemize}
It is clear from Table \ref{tab:Complexity-comparison} that the proposed
algorithms, both linear and nonlinear, reduce the complexity by more
than an order of magnitude when compared to matching pursuits algorithms.
One must of course take into account that while matching pursuits
can estimate the sinusoidal parameters directly from the input signal,
the proposed method requires initial frequency estimates.

\begin{table}
\begin{center}%
\begin{tabular}{ccc}
\hline 
Algorithm & Complexity & Typical (Mflops)\tabularnewline
\hline 
MP (slow) & $4LN^{2}P$ & 3,300\tabularnewline
MP (FFT) & $\frac{5}{2}LNP\log_{2}LP$ & 1,300\tabularnewline
linear (\ref{eq:explicit-least-square}) & $64N^{3}+32LN^{2}$ & 900\tabularnewline
non-linear (\cite{vos50hqc}) & $O\left(N^{4}+LN^{2}\right)$ & >500$^{*}$\tabularnewline
linear (proposed) & $\left(8M+5\right)LN$ & 27\tabularnewline
non-linear (prop.) & $\left(17M-4\right)LN$ & 60\tabularnewline
\hline 
\end{tabular}\end{center}

\caption{Complexity comparison of various parameter estimation algorithms.
$^{*}$The typical complexity of \cite{vos50hqc} is not given, but
we estimate it to be greater than 500 Mflops. \label{tab:Complexity-comparison} }

\end{table}

\section{Conclusion\label{sec:Conclusion}}

We have presented a method for estimating sinusoidal parameters with
very low complexity. Our proposed method is based on a linearisation
of the sinusoidal model, followed by an iterative optimisation of
the parameters. The algorithm converges quickly, in only 2 iterations
for the linear optimisation and 3 iterations for the non-linear optimisation.
It was also shown that the frequency estimation of the non-linear
version of our algorithm is more accurate than the matching pursuits
and time frequency reassignment methods for the experiment. In addition,
we calculated that the complexities of our algorithms were considerably
lower than the matching pursuits algorithms.

Like other non-linear optimisation methods, the method we propose
requires a good initial estimate of the sinusoids' frequencies. Therefore,
low-complexity sinusoid selection is another important problem to
investigate for improving sinusoidal parameter estimation. Also, for
applications that require it, the proposed algorithm could easily
be extended to estimate the frequency modulation within a frame.

\appendices

\section{Linearisation of the Sinusoidal Model\label{sec:Linearisation}}

Let us consider a sinusoidal model with piecewise linear amplitude
modulation and a frequency offset (from an initial estimate):
\begin{multline}
\tilde{x}\left(n\right)=\sum_{k=1}^{N}\left(A_{k}+nA_{k}^{'}\right)\cdot\\
\cos\left(\left(\theta_{k}+\Delta\theta_{k}\right)n+\phi_{k}\right)\ ,
\end{multline}
where $\theta_{k}$ is known in advance and $\Delta\theta_{k}$ is
considered small. Using trigonometric identities, we can expand the
sum in the cosine term as
\begin{align}
\tilde{x}\left(n\right) & =\sum_{k=1}^{N}\left(A_{k}+nA_{k}^{'}\right)\cos\phi_{k}\cos\left(\theta_{k}+\Delta\theta_{k}\right)n\nonumber \\
 & -\sum_{k=1}^{N}\left(A_{k}+nA_{k}^{'}\right)\sin\phi_{k}\sin\left(\theta_{k}+\Delta\theta_{k}\right)n\\
 & =\sum_{k=1}^{N}\left(A_{k}+nA_{k}^{'}\right)\cos\phi_{k}\cos\Delta\theta_{k}n\cos\theta_{k}n\nonumber \\
 & -\sum_{k=1}^{N}\left(A_{k}+nA_{k}^{'}\right)\cos\phi_{k}\sin\Delta\theta_{k}n\sin\theta_{k}n\nonumber \\
 & -\sum_{k=1}^{N}\left(A_{k}+nA_{k}^{'}\right)\sin\phi_{k}\cos\Delta\theta_{k}n\sin\theta_{k}n\nonumber \\
 & -\sum_{k=1}^{N}\left(A_{k}+nA_{k}^{'}\right)\sin\phi_{k}\sin\Delta\theta_{k}n\cos\theta_{k}n\ .\label{eq:4term-sinusoidal-full-expand}
\end{align}

In the linearisation process, we further assume that $\Delta\theta_{k}n\ll1$
and $A_{k}^{'}n\ll A_{k}$, so we can neglect all second order terms
and above. This translates into the following approximations:
\begin{align}
\sin\Delta\theta_{k}n & \approx\Delta\theta_{k}n\ ,\\
\cos\Delta\theta_{k}n & \approx1\ ,\\
nA_{k}^{'}\sin\Delta\theta_{k}n & \approx0\ .
\end{align}
When substituting the above approximations into (\ref{eq:4term-sinusoidal-full-expand}),
we obtain:
\begin{align}
\tilde{x}\left(n\right) & =\sum_{k=1}^{N}\left(A_{k}+nA_{k}^{'}\right)\cos\phi_{k}\cos\theta_{k}n\nonumber \\
 & -\sum_{k=1}^{N}A_{k}\cos\phi_{k}\Delta\theta_{k}n\sin\theta_{k}n\nonumber \\
 & -\sum_{k=1}^{N}\left(A_{k}+nA_{k}^{'}\right)\sin\phi_{k}\sin\theta_{k}n\nonumber \\
 & -\sum_{k=1}^{N}A_{k}\sin\phi_{k}\Delta\theta_{k}n\cos\theta_{k}n\ .\label{eq:4term-sinusoidal-gather}
\end{align}
Reordering the terms in (\ref{eq:4term-sinusoidal-gather}), leads
to the following formulation:
\begin{align}
\tilde{x}\left(n\right) & =\sum_{k=1}^{N}A_{k}\cos\phi_{k}\cos\theta_{k}n\nonumber \\
 & -\sum_{k=1}^{N}A_{k}\sin\phi_{k}\sin\theta_{k}n\nonumber \\
 & +\sum_{k=1}^{N}\left(A_{k}^{'}\cos\phi_{k}-A_{k}\Delta\theta_{k}\sin\phi_{k}\right)n\cos\theta_{k}n\nonumber \\
 & -\sum_{k=1}^{N}\left(A_{k}^{'}\sin\phi_{k}+A_{k}\Delta\theta_{k}\cos\phi_{k}\right)n\sin\theta_{k}n\ ,\label{eq:Taylor-equivalent}
\end{align}
which is a linear combination of four functions. The result in (\ref{eq:Taylor-equivalent})
is in fact equivalent to a first-order Taylor expansion. Keeping second
order terms would allow us to model both the first derivative of the
frequency with respect to time and the second derivative of the amplitude.

\bibliographystyle{IEEEtran}
\bibliography{sinusoids}

\end{document}